\documentclass[a4paper,11pt]{article}
\usepackage{jinstpub} 
\usepackage{subcaption}
\usepackage{siunitx}
\usepackage{graphicx}

\usepackage{adjustbox}
\usepackage{lineno}
\usepackage{float}

\proceeding{iWoRiD 2025\\
 Jul 6–10, 2025\\
Bratislava, Slovakia}

\title{In-situ Radiation Damage Study of Silicon Carbide Detectors Subjected to Clinical Proton Beams}

\newcommand{\peqv}[1]{\mathrm{p}_{#1}^+/\mathrm{cm}^{2}}

\author[1,a]{Daniel Radmanovac,\note{Corresponding author.}}
\author[a]{Andreas Gsponer,}
\author{Simon Waid,}
\author[a]{Sebastian Onder,}
\author[a]{Matthias Knopf,}
\author{Jürgen Burin,}
\author{Stefan Gundacker,}
\author{Thomas Bergauer}

\affiliation{Marietta Blau Institute for Particle Physics, Austrian Academy of Sciences (former HEPHY),\\Dominikanerbastei 16, 1010 Vienna, Austria}
\affiliation[a]{Institute of Atomic and Subatomic Physics, TU Wien, Stadionallee 2, 1020 Vienna, Austria}

\emailAdd{daniel.radmanovac@oeaw.ac.at}

\abstract{
Silicon carbide (SiC) planar PiN diodes from two different manufacturers were irradiated with \SI{252.7}{\mega\electronvolt} protons from a medical synchrotron. Over the course of two 8h irradiation shifts, the samples were exposed to increasing fluences ranging from $1.4\times 10^{11}$ to $3.5\times 10^{13} \; \peqv{}$. Electrical characterizations, including IV and CV measurements, were performed both before and after irradiation using probe stations, and for selected samples even in-situ between fluence steps directly at the irradiation facility. 
The results show a gradual compensation of the effective epitaxial doping concentration with each incremental fluence step, observed as a reduction in capacitance before full depletion and confirmed by the extracted effective doping concentration. From these measurements, linear donor removal rates are determined for all sample groups, with values ranging from $4.2 \mathrm{\;cm^{-1}}$ to $ 6.4\mathrm{\;cm^{-1}}$. These findings provide a quantitative basis for understanding radiation-induced charge carrier removal in 4H-SiC devices and are relevant for predicting the performance and lifetime of future radiation-hard detector technologies, including 4H-SiC LGADs.
}


\arxivnumber{2510.11304} 

\begin{document}
\maketitle
\flushbottom

\section{Introduction}
\label{sec:intro}

\subsection{Silicon carbide as detector material}

Silicon carbide (SiC) is a wide-bandgap semiconductor known for its high breakdown voltage, charge carrier drift velocity, and thermal conductivity, making it widely used in power electronics and a promising alternative to silicon in high-energy physics applications. The most commonly used polytype (crystal stacking order) in industry is 4H-SiC, which has a bandgap energy of $E_\mathrm{g} = \SI{3.26}{\electronvolt}$ \cite{sinelnik_theoretical_2021}. Its ten-fold higher breakdown voltage over silicon allows power electronics devices to be made ten times as thin, which means an order of magnitude faster charge carrier drift time. In 4H-SiC detectors, the wide bandgap enables operation at room temperature even after significant irradiation, due to the low dark current.

One of the simplest forms of a semiconductor particle detector is a reverse-biased p-in-n (PiN) diode. On a highly-doped 4H-SiC substrate wafer, typically around \SI{300}{\micro\meter} thick, a high-purity epitaxial layer (epi-layer) with a thickness of up to $\SI{100}{\micro\meter}$ is grown. The minimum doping of the epitaxial layer that can be achieved is around \SI{5e13}{\centi\meter^{-3}}, and is still the focus of optimization in the manufacturing process. On top of the epi-layer, a thin, strongly p-doped layer is implanted, and both anode and cathode metal contacts are deposited on top and bottom of the device. When a reverse bias is applied, the lightly doped epi-layer gets depleted, forming the active volume of the detector. Further details about the fabrication process including ion implantation can be found in \cite{novotny_first_2025} and \cite{rodriguez_planar_2005}. 

Due to its larger ionization energy compared to silicon, 4H-SiC produces fewer charge carriers (about \num{55} to \num{57} electron–hole pairs per \si{\micro\meter} for a minimum ionizing particle (MIP) \cite{gsponer_measurement_2024}). Increasing the thickness of the active volume would enhance the signal amplitude, but this is challenging to manufacture and leads to very large full depletion voltages due to the limitations on the doping concentrations that can be obtained, as the depletion width grows proportional to the square root of the applied bias voltage and inverse doping concentration.
\begin{equation}
\label{eqn:depl_width}
    w = \sqrt{\frac{2\varepsilon_0 \varepsilon_\mathrm{SiC} ( V_\mathrm{bi} + V_\mathrm{bias})}{eN_\mathrm{eff}}}.
\end{equation}

To overcome the challenge of small signal amplitudes and to increase the signal-to-noise ratio, Low-Gain Avalanche Detectors (LGADs) have become increasingly popular in recent years. LGADs incorporate a carefully engineered, highly doped gain layer that creates a localized high-electric-field region near the junction. This region enables controlled avalanche multiplication, amplifying the initial charge signal directly inside the sensor before it reaches the readout electronics. This built-in gain improves timing resolution and signal-to-noise ratio, making LGADs particularly well-suited for applications such as particle tracking and precision timing in pile-up-prone high-luminosity experiments. The amplification of an LGAD is an essential puzzle piece for the application of 4H-SiC detectors in high-energy physics, and several groups are in the process of developing and testing such devices \cite{svihra_exploring_2025, novotny_first_2025, yang_ultra-fast_2025, YANG2026170873, ONDER_2025}.
However, the gain in the multiplication layer is highly sensitive to its precise doping concentration, and even small variations or radiation-induced defects might significantly alter the electric field, leading to fluctuations or degradation of the detector’s amplification performance over time. 

\subsection{Radiation damage in silicon carbide}
\label{sec:intro_raddmg}
Critical to the performance of SiC detectors and electronics is their response to radiation damage-induced defects. This is of particular relevance since the trend for high-energy particle physics experiments is moving towards high luminosities, with the imminent high luminosity (HL) upgrade for the LHC and the construction of the FCC in the future \cite{cern_yellow_report_hi_lumi, bartmann_future_2025, the_fcc_collaboration_fcc-hh_2019, hartmann_evolution_2017}. As an example: The expected irradiation fluence received by the innermost detectors in the CMS experiments at CERN during the planned HL-LHC running phase is \SI{2e16}{\centi\meter^{-2}} of \SI{1}{\mega\electronvolt} neutron equivalent n$_\mathrm{eq.}$\cite{CMSSteinbrueck2024}.

In two recent irradiation studies by the authors, a collection of samples were irradiated in TRIGA research reactors in Vienna and Ljubljana up to $1\times 10^{16}\:\text{n}_{\text{eq.}}/\text{cm}^2$ and $1\times 10^{18}\:\text{n}_{\text{eq.}}/\text{cm}^2$ respectively \cite{gsponer_neutron_2023}. The electrical characterization of those irradiated samples showcased interesting results: Although the reverse current in current-voltage (IV) measurements remained below $10\;\si{\pico\ampere}$ for all fluences up to $1\times 10^{16}\:\text{n}_{\text{eq.}}/\text{cm}^2$, the forward current was significantly reduced with increasing fluence and a forward breakdown was only observed within the sample with the lowest irradiation fluence of $5 \times 10^{14}\:\text{n}_{\text{eq.}}/\si{\centi\meter}^2$. In addition, it was also possible to operate the devices in forward bias. This effect has been reproduced in TCAD simulations \cite{burin_tcad_2025} since then and is attributed to an electric field barrier created by space charges, which block electrons and holes from entering the diode. 
 
Through irradiation, deep acceptor-like defects are created, reducing the amount of available free charge carriers in the epitaxial layer. In a recent neutron irradiation study \cite{park_neutron_2023}, the authors were able to quantify a significant increase in Z$_{1/2}$ deep-level traps with deep-level transient spectroscopy (DLTS) measurements. This results in a decrease in measurable effective doping concentration, and the observed capacitance before full depletion decreases accordingly. This is a manifestation of the deep defect level, as those defects appear invisible to regular AC probing frequencies at room temperature. In \cite{gsponer_neutron_2023} and \cite{moscatelli_radiation_2006}, the capacitance was independent of the applied bias voltage and equal to the full depletion capacitance for all irradiation fluences, as no free charge carriers were left in the epitaxial layer.

Little data exists on the decrease of free charge carrier concentration in detail, as most studies aim at higher irradiation fluences, where these processes have already taken course, and the doping of the epitaxial layer is fully compensated. We present a radiation study involving lower irradiation fluences between $1.4\times 10^{11}$ and $3.5\times 10^{13} \; \peqv{}$, to characterize the early effects of radiation damage in silicon carbide, including the rate of donor removal in the epitaxial layer.
This study consists of two 8h irradiation shifts at the MedAustron proton synchrotron with SiC samples from two different manufacturers. In the first shift, two samples were alternately electrically characterized and irradiated in multiple steps. This allowed for the observation of radiation-damage-induced ageing effects at different stages within the same samples.
In the second shift, seven PiN samples have been irradiated "the traditional way", where samples receive their total intended fluence at once and are electrically characterized in the laboratory afterwards.
The results of these irradiation studies and a comparison between PiN samples are presented in the next sections. 

\section{Materials and methods} 
\label{sec:methods}
Planar 4H-SiC samples from two different manufacturers were investigated in this study: two samples from IMB CNM, CSIC (Spain)~\cite{rafi_electron_2020}, and seven samples produced by onsemi (Czech Republic)~\cite{novotny_first_2025, svihra_exploring_2025}.
All samples had an active area of $3\times 3 \; \text{mm}^2$ and an epitaxial thickness of \SI{50}{\micro\meter}
except for one CNM sample (W4) with an epi thickness of \SI{100}{\micro\meter}. Also, the doping concentration, extracted from CV measurements, in the epitaxial layer of this W4 sample is approximately four times larger than the sample from wafer 2 (W2) or the onsemi counterparts. The two different CNM samples will be further referenced as W2 and W4 throughout this study. An overview of all PiN samples featured in this study can be found in Table \ref{tab:fluences}.

All samples underwent electric characterization in the laboratory before irradiation, consisting of IV and CV measurements in a probe station. IV measurements were performed with a Keithley 2657A source measurement unit (SMU) and a 6517B electrometer. CV measurements were performed with a Keysight E4980A LCR meter and a DC bias decoupling box. Both the in-situ CV-measurements (CNM samples) and measurements in probe stations were conducted using a \SI{1}{\mega\hertz} AC frequency and \SI{500}{\milli\volt} AC amplitude.
For unirradiated and irradiated samples, the measured capacitance showed no significant dependence on the frequency.

\begin{table}[H] 
\caption{All irradiated PIN samples with their epi-layer thicknesses and doping concentrations (from CV measurements), and their maximum proton fluences. Sample names starting with 'W' were manufactured by CNM, and the remaining samples by onsemi, respectively.}
\resizebox{\textwidth}{!}{%
    \centering
    \begin{tabular}{l|ccccccccc}
        \hline
        Sample name & W2 & W4 & PIN2 & PIN3 & PIN4 & PIN5 & PIN9 & PIN8 & \\ \hline
        Thickness (\si{\micro\meter}) & 50 & 100 & 50 & 50 & 50 & 50 & 50 & 50 & \\ \hline
        Epi-layer doping (\si{\centi\meter^{-3}}) & $\sim(2-10)\times10^{13}$ & $4\times10^{14}$ & $4\times10^{13}$ & $4\times10^{13}$ & $4\times10^{13}$ & $4\times10^{13}$ & $4\times10^{13}$ & $4\times10^{13}$ & \\ \hline
        
        Fluence ($\peqv{})$ & $2.4\times10^{13}$ & $3.4\times10^{13}$ & $1.4\times10^{11}$ & $6.9\times10^{11}$ & $1.4\times10^{12}$ & $2.8\times10^{12}$ & $6.9\times10^{13}$  & $1.4\times10^{13}$ & \\ \hline

    \end{tabular}
    }
    \label{tab:fluences}
\end{table}

\subsection{MedAustron ion therapy center}

MedAustron is a synchrotron-based ion therapy center situated in Wiener Neustadt (Austria), with one beamline dedicated to research purposes. The synchrotron is capable of accelerating protons between \qty{62.4}{MeV} and \qty{800}{MeV} and carbon ions between \qty{120}{MeV/u} and \qty{400}{MeV/u}. In clinical mode, particle rates up to $\sim 10^{10}\: \text{p}^+$/s and $10^{8}\: \text{C}^{6+}$/s are available, and the delivered dose (fluence) is monitored and controlled with ionization chambers integrated into the beamline.

According to non-ionizing energy loss (NIEL) data in silicon \cite{lindstrom_radiation_2003}, lower energy protons have a higher radiation hardness factor and therefore produce a larger amount of $1\; \text{MeV}$ neutron equivalent radiation damage. However, at \SI{62.4}{\mega\electronvolt}, the beam spot is significantly larger than for higher beam energies. Therefore, we used a \SI{252.7}{\mega\electronvolt} proton beam ($k_{252\:\text{MeV p}^+}\sim 0.9 $) as we estimated to deliver the highest NIEL dose to the samples with this setting.
For a beamtime of 8 hours, this resulted in a maximum achievable proton fluence of around $1\times10^{14}$ $\peqv{}$ on a single detector, equivalent to $9\times10^{13}$ n$_\mathrm{eq.}/\si{\centi\meter}^2$ in silicon.
Over the area of the diodes, the fluence non-uniformity was estimated to be less than \SI{10}{\percent}.

\subsection{In-situ irradiation and characterization} 

\begin{figure}[htbp]  
    \centering
    \begin{minipage}[b]{0.25\textwidth}
        \includegraphics[width=\linewidth]{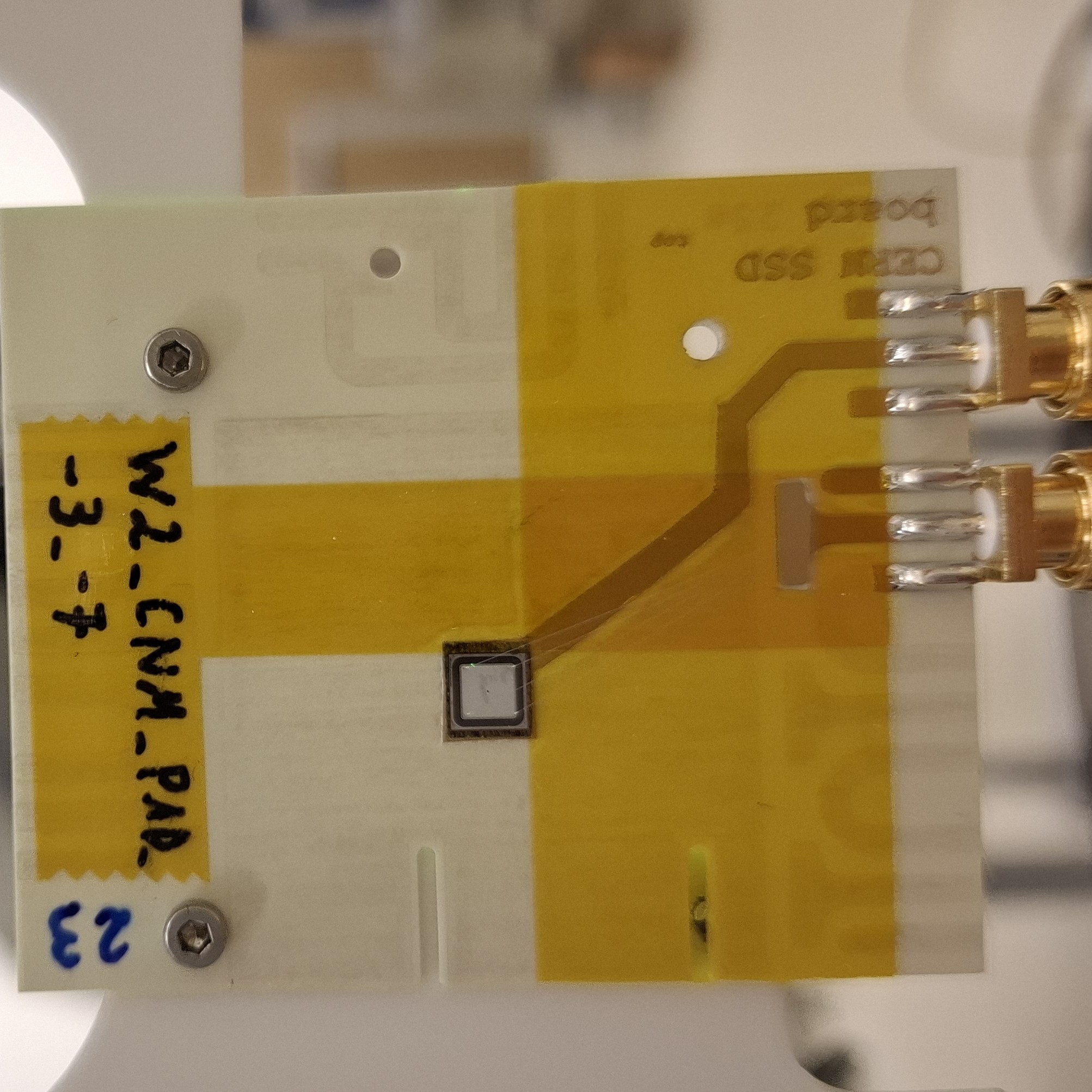}
    \end{minipage}
    \begin{minipage}[b]{0.475\textwidth}
        \includegraphics[width=\linewidth]{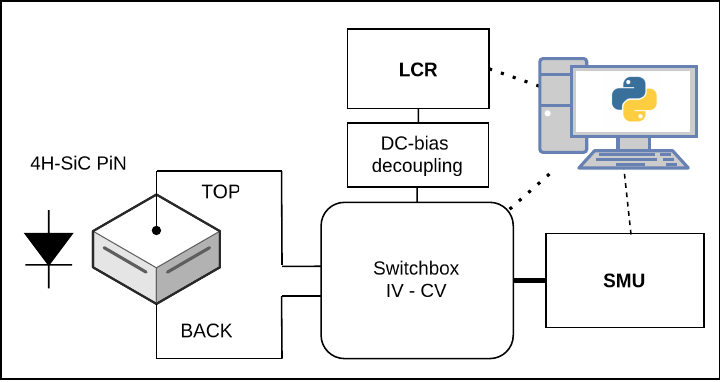}
    \end{minipage}
    \begin{minipage}[b]{0.25\textwidth}
        \includegraphics[width=\linewidth]{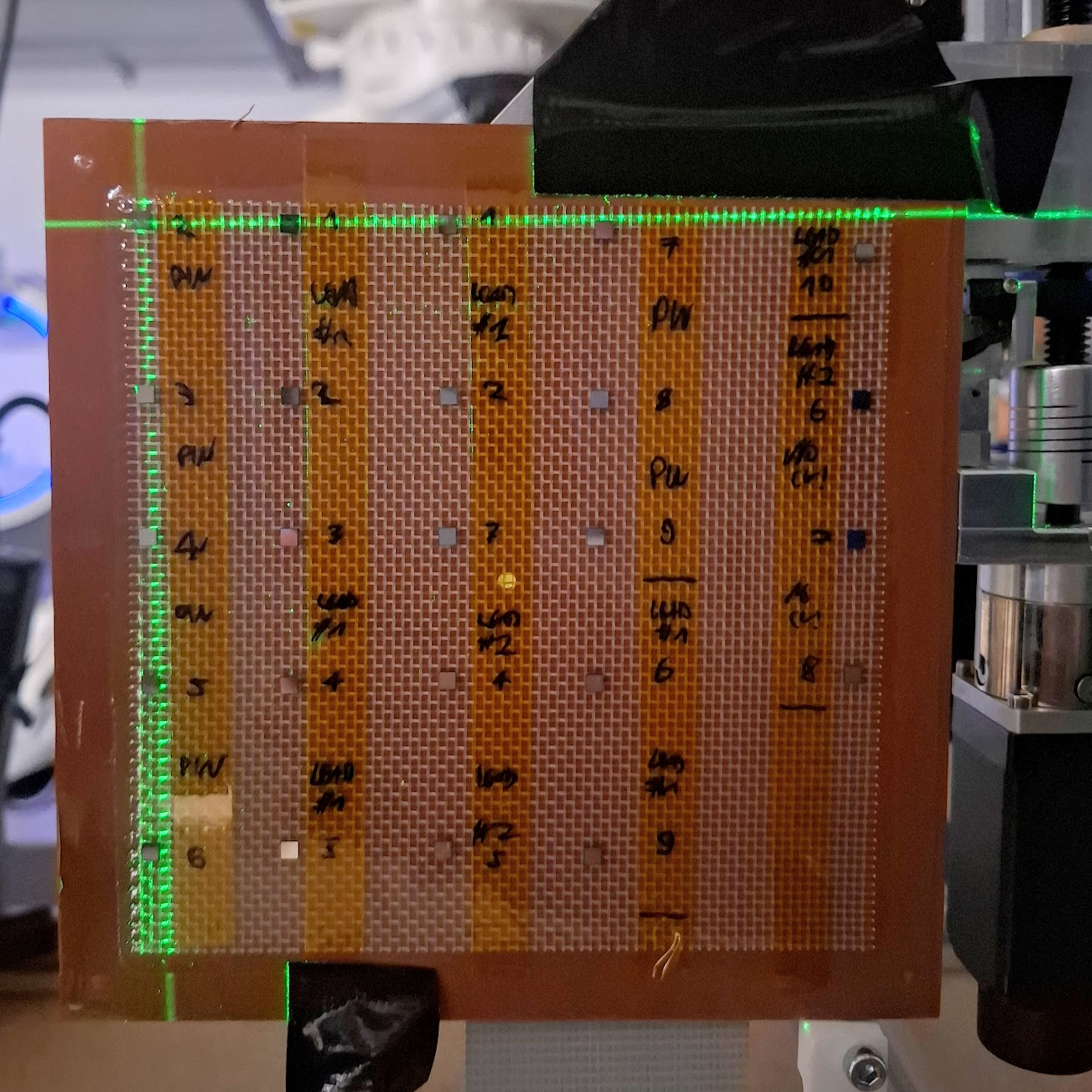}
    \end{minipage}
    
    \caption{Left: CNM SiC-diode mounted on a ceramic board with bond-wires connecting the top-side to a separate connector. SMA cables are connected to a switchbox during the in-situ proton irradiation. Diagram of the in-situ setup in the middle. Right: Gel-pak with SiC-detector samples mounted in front of the beam nozzle. The beam's isocenter is located at the crossing point of all three alignment lasers.}
    \label{fig:ir1_setup}
\end{figure}

Due to limited sample availability and beam time, we adopted a novel approach for the first irradiation shift. Both CNM 4H-SiC PiN diodes were mounted on separate ceramic boards, with their top and back sides connected to two separate channels of a switchbox with triaxial relays.
The switchbox was connected to a Keithley 2470 SMU and an Agilent 4284A LCR meter, together with a DC-bias decoupling box. A custom Python script enabled remote control of the switchbox and automated the execution of I–V and C–V measurements. A picture of the mounted diode and a diagram of the measurement setup can be seen in Figure \ref{fig:ir1_setup} (a,b).
At the start of the shift and after each irradiation step, a full measurement cycle was performed. This included forward and reverse I–V sweeps, as well as two C–V measurements: one coarse scan up to \qty{750}{V} with a larger step size, and a fine scan up to \qty{10}{V} with small step sizes to capture the rapidly changing initial region of the C–V curve in greater detail.
To balance measurement accuracy with limited shift time, six irradiation steps were performed, each followed by a measurement cycle lasting approximately 25 minutes. The two samples were irradiated and measured sequentially, with roughly half the shift time ($\sim 4\:$h) dedicated to each sample.\\

In the second shift, the in-situ measurement setup was replaced to accommodate a larger number of samples, of which seven PiN diodes are featured in this study. The samples were arranged and fixed on a $25\times 25\, \text{cm}^2$ Gel-Pak carrier plate and spaced approximately \qty{5}{cm} apart, as shown in Figure \ref{fig:ir1_setup} (c). The carrier plate was mounted on aluminium clamps connected to a stepper motor, enabling automated vertical and horizontal movement.
Each diode was irradiated individually by positioning it at the beam’s iso-center using the stepper motor. 
Treatment plans specifying an exact number of particles were used to irradiate the individual samples.
No significant activation of the samples was detectable two weeks after the irradiation, and all samples were electrically characterized in the lab again. For a list of all samples and their received total fluences see Table \ref{tab:fluences}.

\section{Results}
\label{sec:results}
\subsection{Reduction in forward current}
\begin{figure}[htbp]

    \begin{minipage}[b]{0.33\linewidth}
        \includegraphics[width=\linewidth,
        trim={0.25cm 0 0 0},clip]{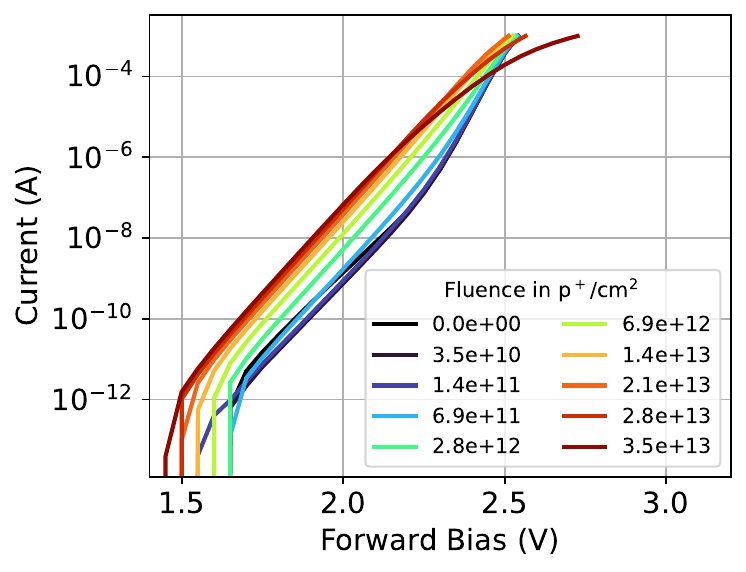}
       \centering (a) CNM W4
    \end{minipage} 
    \begin{minipage}[b]{0.33\linewidth}
        \includegraphics[width=\linewidth, trim={0.25cm 0 0 0},clip]{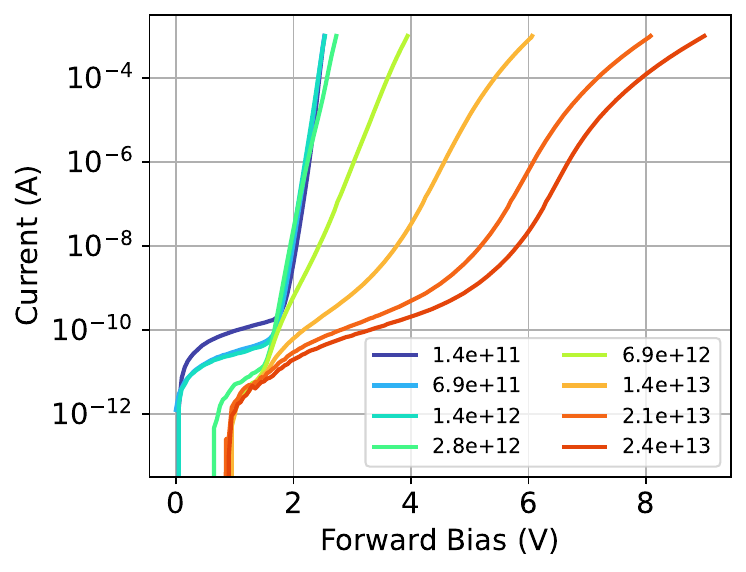}
        \centering (b) CNM W2
    \end{minipage}
    \begin{minipage}[b]{0.33\linewidth}
        \includegraphics[width=\linewidth, trim={0.25cm 0 0 0},clip]{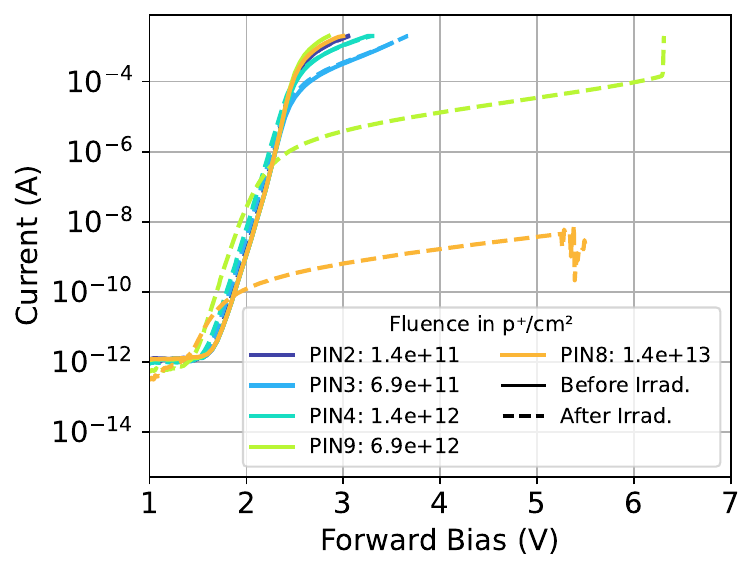}
        \centering (c) onsemi PiNs
    \end{minipage} 
    
    \caption{Forward I-V in-situ measurement results of two planar CNM PiN samples from different wafers (a,b) and probe station I-V measurements on irradiated onsemi samples (c). The epitaxial doping concentration in the W4 sample (a) is higher, and hardly any change in forward current is visible throughout the applied fluence range. With lower epitaxial doping, we can observe that the exponential increase in forward current is delayed to higher bias voltages with each irradiation step due to the growth of an electric field barrier near the electrodes.
    }
    \label{fig:iv-fw_cnm}
\end{figure}

While there was no significant change in reverse current after the irradiation in all samples, a significant fluence-dependent suppression in forward current was observable in samples with lighter epitaxial doping (CNM W2 and onsemi samples). The results for the two CNM and the onsemi PiN samples are presented in Figure \ref{fig:iv-fw_cnm}. The forward current becomes increasingly suppressed, and the exponential increase in current is delayed at larger fluences. 
This effect was not observable in the sample with a high epitaxial doping (CNM W4, Figure \ref{fig:iv-fw_cnm}a), which indicates that for higher-doped samples, more fluence is necessary for the same results.

\subsection{Capacitance decrease before full depletion}
\begin{figure}[htp]
        
    \begin{minipage}[b]{0.33\linewidth}
        \includegraphics[width=\linewidth, trim={0.25cm 0 0 0},clip]{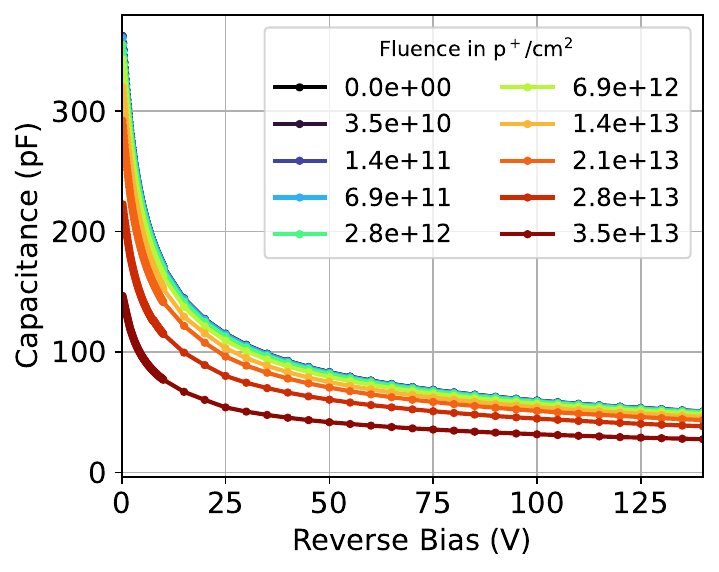}
            \end{minipage}
    \begin{minipage}[b]{0.325\linewidth}
        \includegraphics[width=\linewidth, trim={0.25cm 0 0 0},clip]{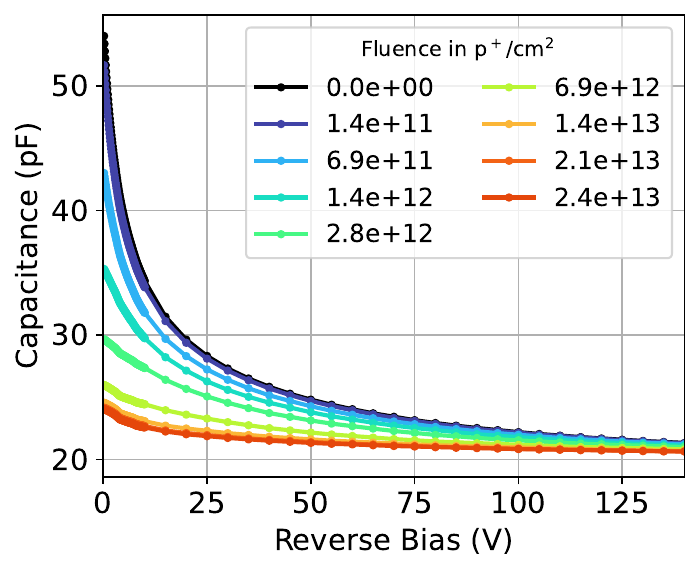}
        \end{minipage}
    \begin{minipage}[b]{0.33\linewidth}
        \includegraphics[width=\linewidth, trim={0.25cm 0 0 0},clip]{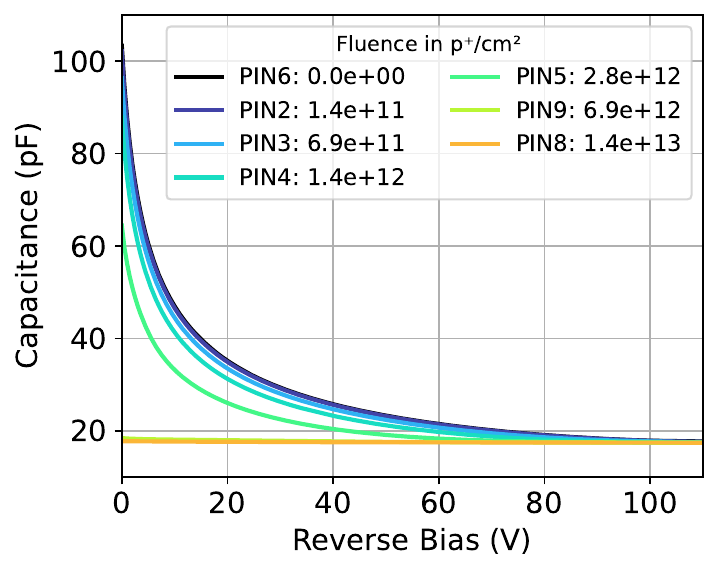}
            \end{minipage}

    \:\:\:\begin{minipage}[b]{0.32\linewidth}
        \includegraphics[width=\linewidth, trim={0.25cm 0 0 0},clip]{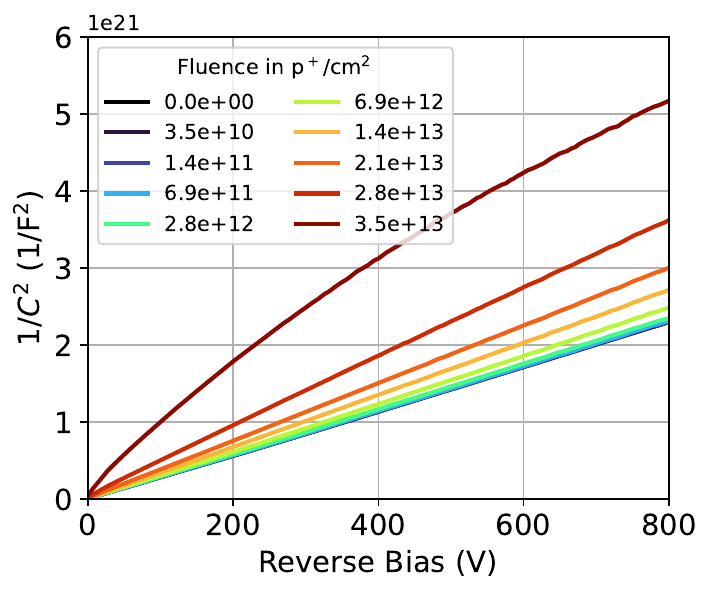}
        \centering (a) CNM W4
    \end{minipage}
    \begin{minipage}[b]{0.33\linewidth}
        \includegraphics[width=\linewidth, trim={0.25cm 0 0.1 0},clip]{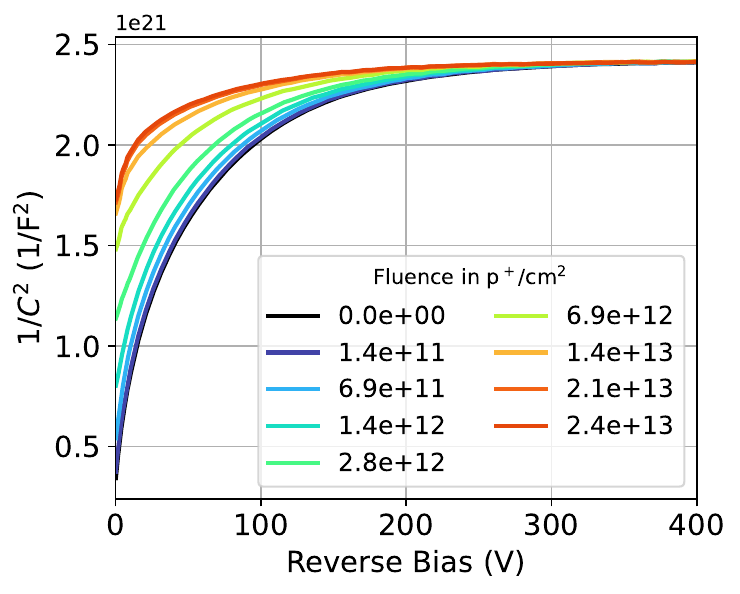}
        \centering (b) CNM W2
    \end{minipage}
    \begin{minipage}[b]{0.325\linewidth}
        \includegraphics[width=\linewidth, trim={0.2cm 0 0.25cm 0},clip]{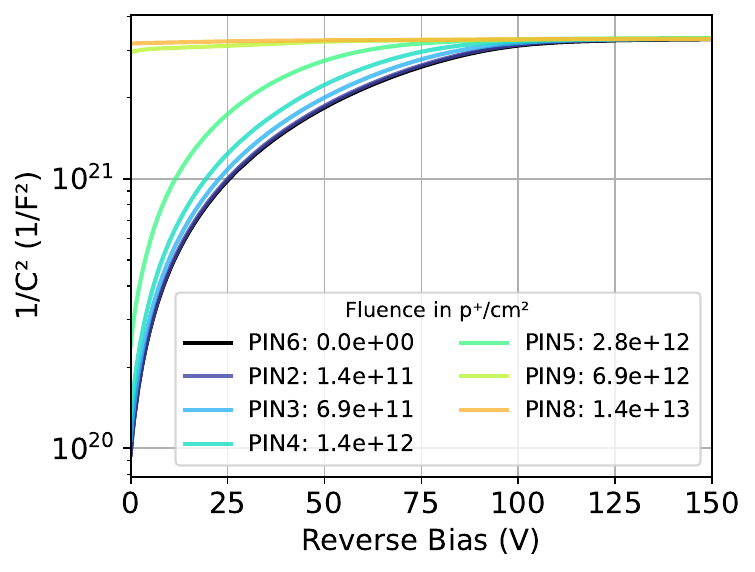}
        \centering (c) onsemi PiNs
    \end{minipage}
    \caption{CV-measurements of all samples zoomed on lower voltages. In both the in-situ (a,b) and traditional irradiation campaign (c), samples show a gradual reduction in the measured capacitance before full depletion. 
    }
    \label{fig:cv_zoom_all}
\end{figure}

By starting with small irradiation steps of $\sim 1\times 10^{11}\:\peqv{}$, it was possible to measure and visualize the gradual decrease in capacitance upon irradiation. 
 The results of the C-V-measurements focused on lower bias voltages are presented in Figure \ref{fig:cv_zoom_all}. In lightly doped samples (b,c), it was possible to reproduce the flat-lining of the capacitance observed in high-fluence studies~\cite{gsponer_neutron_2023}, while also showcasing the gradual transformation through intermediate steps. In the higher-doped sample (a), constant capacitance was not achieved with the applied fluence, although a clear trend is visible. 
 As described in section \ref{sec:intro_raddmg}, the observed decrease in capacitance before full depletion is a manifestation of dopant charge carriers in the epitaxial layer being trapped in acceptor like traps introduced to the material by irradiation. A constant CV-curve at all reverse bias voltages suggests a fully compensated material, as no free dopant charge carriers are available any more.

\subsection{Donor removal rate}

The change of the effective doping in the depletion area over low radiation fluences can be described with the following linear expression \cite{moll_displacement_2018, ruddy_review_2024}:
\begin{equation}
\label{eqn:neff_drr_lin}
 N_\mathrm{eff}(\Phi) = N_0 - g\Phi.
\end{equation}
$N_0$ and $N_\mathrm{eff}$ are the effective epitaxial doping concentrations before and after irradiation with the fluence $\Phi$ (\si{\centi\meter}$^{-2}$). The linear donor or carrier removal rate $g$ (\si{\centi\meter}$^{-1}$) is dependent on the radiation type and energy and corresponds in our case to protons with an energy of \SI{252.7}{\mega\electronvolt}. As mentioned before, according to the NIEL hypothesis, this can be put in comparison with \SI{1}{\mega\electronvolt} neutrons in silicon, by multiplying it with the radiation hardness factor $k$ for \SI{252.7}{\mega\electronvolt} protons. Swapping the \SI{252.7}{\mega\electronvolt} protons with \SI{1}{\mega\electronvolt} neutrons would therefore yield fluences $\sim10\%$ below our reported values.

\begin{figure}[htp]
        \begin{minipage}[b]{0.505\textwidth}
        \includegraphics[width=\linewidth,trim={0.25cm 0 0 0}, clip]{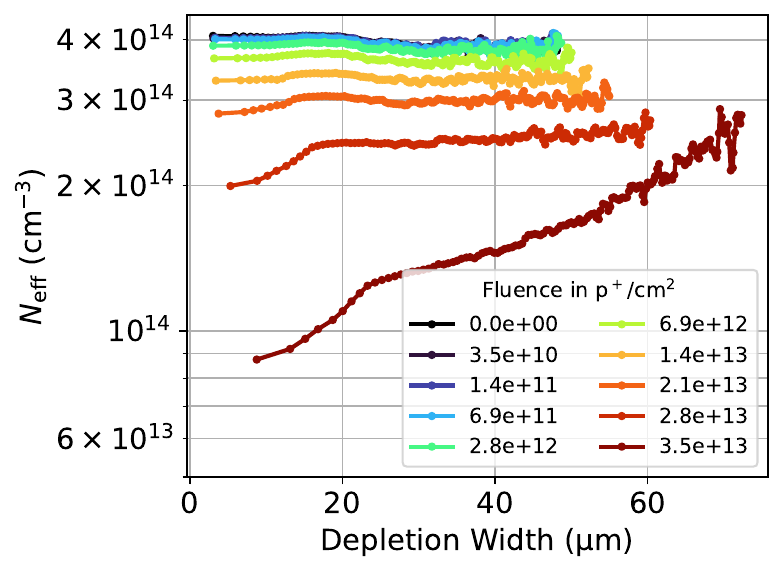}
        \centering (a) CNM W4
    \end{minipage}
    \begin{minipage}[b]{0.48\linewidth} 
        \includegraphics[width=\linewidth,trim={0.25cm 0 0 0}, clip]{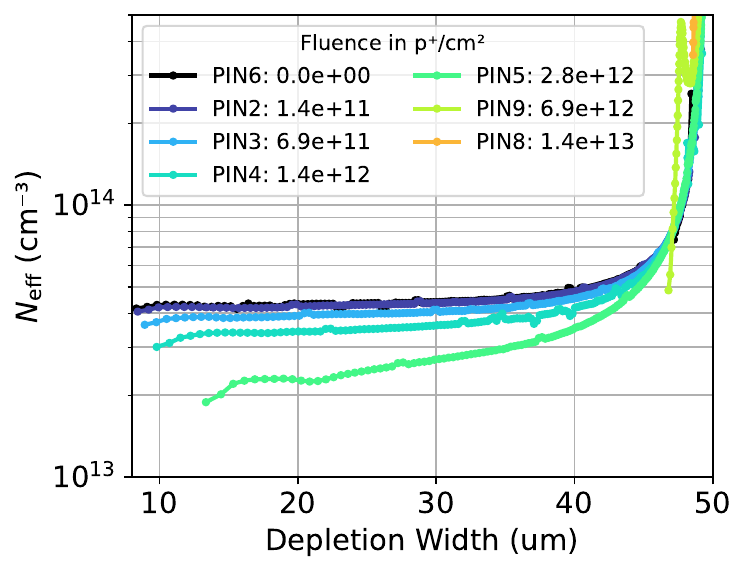}
        \centering (b) onsemi PiNs
    \end{minipage}
    
    \caption{Doping profiles calculated from CV-measurements for the CNM W4 sample (a) and onsemi PiN samples (b). CV-data in (a) was smoothed using a low-pass filter before the calculation of the doping concentration, as the measured data was noisier in the in-situ measurement.
    }
    \label{fig:doping_profiles}
\end{figure}

By comparing the effective doping concentration $N_\mathrm{eff}$ of the same sample (W2, W4) or of similar samples (onsemi PiNs) at different amounts of applied fluence $\Phi_\mathrm{p^+}$, we can verify the linear relationship from \eqref{eqn:neff_drr_lin} and obtain the donor removal rate $g$ as a fit parameter. For strongly uniform doping concentrations throughout the epitaxial width, the measured $1/C^2$ is linear over bias voltage before full depletion, and the slope of this line can be used to calculate $N_\mathrm{eff}$ for the given sample and fluence.
To avoid linearisation errors, as most samples showed non-linear behaviour in $1/C^2$, we chose to compare the effective doping concentrations at similar depletion widths.
In Figure \ref{fig:doping_profiles}, the extracted doping profiles for the CNM W4 sample and onsemi PiN samples are presented. The doping profile of the W2 sample is not shown due to higher noise levels, but it exhibits similar overall behaviour with a less linear trend.

The change between moderately decreasing capacitance to nearly constant behaviour happens rapidly, which also manifests in a jump in calculated depletion widths. 
Lightly doped samples only provide data points at low fluences, since the calculated depletion width very quickly approaches the device thickness, as observable in Figure \ref{fig:doping_profiles}b for the highest two fluences. 
The strongly doped W4 sample thus provides more data points at higher fluences for the estimation of the donor removal rate.

\begin{figure}
    \centering
    \begin{minipage}[b]{0.65\linewidth}
        \includegraphics[width=\linewidth]{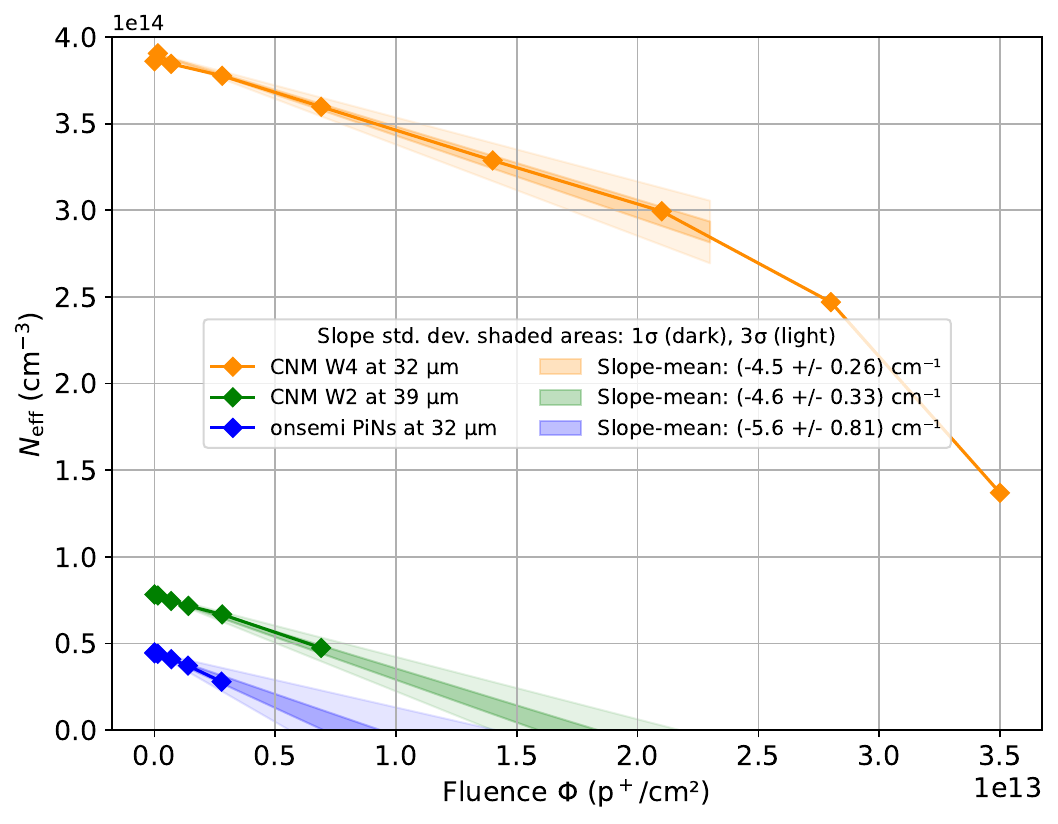}
    \end{minipage}
    \begin{minipage}[b]{0.255\linewidth}
        \includegraphics[width=\linewidth]{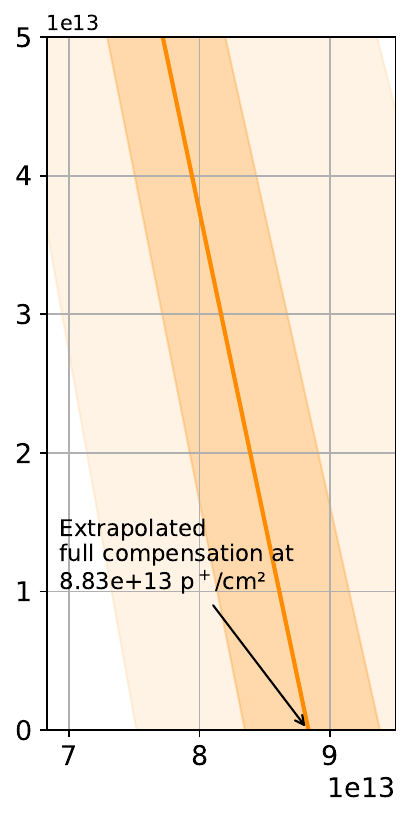}
    \end{minipage}
    \caption{
    Donor removal rate fitted for all samples, and discrepancy between chosen common depletion widths shown as uncertainty as shaded areas (one/three standard-deviations). Due to the high epi-doping of the CNM W4 sample, full compensation was not achieved, and a linear fluence projection until full dopant compensation according to the line fit is shown on the right hand side.
    }
    \label{fig:neff_vs_fluence_all}
\end{figure}

In Figure \ref{fig:neff_vs_fluence_all}, we present the summarized results of the changes in effective doping concentration for all tested samples. Linear fits of the donor removal rate have been performed at each depletion width in the ranges shown in Table \ref{tab:drr}. One representative fit for each sample is drawn with their corresponding data points, while the average of those fits is drawn as a similarly colored area. Outliers where the fit did not yield meaningful results due to noisy data were removed. A standard error of $0.2\; \mathrm{cm^{-1}}$ in the slope of the line fit was chosen as the cut-off. 
\begin{table}[H]
\centering
\caption{Calculated donor removal rates $g$ for both in-situ irradiated CNM samples and for the irradiated onsemi PiN samples.}
\resizebox{0.7\textwidth}{!}{%
    \begin{tabular}{l|l|c|c}
    \hline
    Sample & $g (\mathrm{cm^{-1}})$ & depletion width range $(\mathrm{\upmu m})$ & number of fits \\ \hline
    CNM W4 & \SI[separate-uncertainty=true]{4.48(0.26)}{\centi\meter^{-1}} & 10 - 40 & 230 \\
    CNM W2 &\SI[separate-uncertainty=true]{4.6(0.33)}{\centi\meter^{-1}} & 30 - 40 & 20 \\
    onsemi & \SI[separate-uncertainty=true]{5.6(0.81)}{\centi\meter^{-1}} & 10 - 40 & 150 \\ \hline
    \end{tabular}
    }

\label{tab:drr}
\end{table}
Properly extracting the data of the CNM W2 sample posed a challenge, as the in-situ CV data was very noisy, especially towards larger depletion depths. This is reflected in the relatively low total amount of line fits at different depletion depths (20).
Due to its higher epi-layer doping concentration, the CNM W4 sample provides data points at higher fluence levels. Exceeding $2\times 10^{13}\; \peqv{}$, a non-linear behaviour is emerging at all common depletion widths. The linear fits for this sample therefore exclude the two highest fluences. Extrapolating from the linear region of donor removal indicates full compensation at a fluence of approximately $\sim 8.8\times10^{13} \peqv{}$.

\section{Conclusion}
\label{sec:discussion}

In this study, we presented the results of two low-fluence proton irradiation campaigns conducted at an ion therapy center, using samples from two different manufacturers. These experiments demonstrated a gradual reduction in measured capacitance as the free charge carriers of the epitaxial doping are caught in deep-level traps. From the reduction of the measured capacitance, the donor removal rate was quantified, yielding results in the range of $4.2 \mathrm{\;cm^{-1}} -6.4\mathrm{\;cm^{-1}}$ for all three groups of samples. Such rates are particularly relevant for predicting the operational lifetime of 4H-SiC LGAD devices, as the device gain is highly sensitive to the doping of the gain layer. First irradiation studies with 4H-SiC LGADs \cite{satapathy_impact_2025, zhao_study_2025} indicate significant gain loss upon irradiation. While this work focused on the measured changes of active doping concentration in SiC-PiN samples, a follow-up study on the consequences for signal generation is planned. Early test indicate a change in charge collection efficiency, but further analysis is required. The results presented in this work also provide a foundation and clear direction for future studies, which should focus on further investigations on irradiated LGAD samples and on the radiation-induced changes of the gain layer, to improve our understanding of the underlying processes and develop potential methods to compensate them. 

The first irradiation campaign employed an in-situ approach, where the same sample was iteratively irradiated and characterized. This methodology minimizes uncertainties introduced by sample-to-sample variations, allowing the effects of radiation damage to be studied with high internal consistency. However, it comes with notable drawbacks: measurements cannot be repeated once the sample has been irradiated to higher fluences, and noise levels at the beam line were considerably higher than at a probe station in the clean room. Additionally, time constraints on the beamtime limit the available time and step size that can be used for the measurements.

Finally, our results highlight that proton synchrotrons at ion therapy centers offer a suitable environment for low-fluence irradiation studies. Their well-characterized beam parameters and precise dose delivery make them an excellent choice for controlled investigations of radiation effects in semiconductors.

\acknowledgments

This work was supported by the Austrian Research Promotion Agency (FFG), grant number 916317.
The financial support of the Austrian Ministry of Education, Science and Research is gratefully acknowledged for providing beam time and research infrastructure at MedAustron.

\bibliographystyle{JHEP}
\bibliography{references.bib}

\end{document}